\documentclass[12pt,preprint]{aastex}
\usepackage{emulateapj5}

\shorttitle{60 kpc Optical Filament in Coma Cluster}
\shortauthors{Yagi et al.}

\def\Ha{H$\alpha $}
\def\Hb{H$\beta $}

\def\NB{NB}

\begin{document}

\title{
The Remarkable 60$\times$2 kpc Optical Filament 
Associated with a Poststarburst Galaxy
in Coma Cluster\footnotemark[1]}
\footnotetext[1]{Based on data collected at the Subaru Telescope, 
which is operated by the National Astronomical Observatory of Japan.}

\author{
Masafumi Yagi\altaffilmark{2,3},
Yutaka Komiyama\altaffilmark{3,4},
Michitoshi Yoshida\altaffilmark{5},
Hisanori Furusawa\altaffilmark{4},
Nobunari Kashikawa\altaffilmark{3},
Yusei Koyama\altaffilmark{6},
Sadanori Okamura\altaffilmark{6,7}}

\altaffiltext{2}{email:yagi@optik.mtk.nao.ac.jp}
\altaffiltext{3}{
Optical and Infrared Astronomy Division,
National Astronomical Observatory of Japan,
Mitaka, Tokyo, 181-8588, Japan}
\altaffiltext{4}{Subaru Observatory,
National Astronomical Observatory of Japan,
650 North A'ohoku Place
Hilo, HI 96720, U.S.A.}
\altaffiltext{5}{Okayama Astrophysical Observatory,
National Astronomical Observatory of Japan,
Kamogata, Okayama 719-0232, Japan}
\altaffiltext{6}{Department of Astronomy, School of Science,
University of Tokyo, Bunkyo-ku, Tokyo 113-0033, Japan}
\altaffiltext{7}{Research Center for the Early Universe, 
University of Tokyo,  7-3-1 Hongo, Bunkyo-ku, Tokyo 113-0033, Japan}

\begin{abstract}
In the deep narrow band image of the Coma Cluster taken with
Suprime-Cam of the Subaru telescope, we found an extremely long and
narrow ($\sim$60 kpc$\times$2 kpc) \Ha\ emitting region associated
with a poststarburst galaxy (D100).  Follow up spectroscopy shows that
the region has the same redshift as D100.  The surface brightness of
the region is typically 25 mag(AB) arcsec$^{-2}$ in \Ha\, which
corresponds to 0.5 -- 4$\times$10$^{-17}$erg s$^{-1}$ cm$^{-2}$
arcsec$^{-2}$.  We set two possible explanations for the origin of the
region; gas stripped off from a merged dwarf, or gas stripped off from
D100 by ram pressure.  Either scenario has difficulty to fully explain
all the observed characteristics of the region.
\end{abstract}


\section{Introduction}

Deep H-alpha imaging provides an opportunity to detect new faint
extended features, even in well studied regions
\citep[e.g.,][]{Veilleux2003}.  In nearby cluster Abell 1367,
\citet{Gavazzi2001} found extended ionized regions associated with
starburst Irr galaxies.  The size of the regions are 75$\times$8 kpc,
and 50$\times$8 kpc.  \citet{Yoshida2002} discovered another extremely
extended ionized region ($\sim$35 kpc) in Virgo cluster, which is
found to be a part of 110$\times$25 kpc HI gas \citep{Oosterloo2005}.
\citet{Veilleux2003} used tunable filter for searching ionized gas
around nearby galaxies.  They found $\sim$ 20 kpc complexes/filaments
in 6 galaxies out of 10.  Another type of elongated gas is Magellanic
Stream \citep[hereafter MS;][]{Mathewson1974}.  MS is orbiting around
our Galaxy, and emitting \Ha\ \citep[hereafter WW96]{Weiner1996}.  The
connection between such ionized region and the evolution of associated
galaxy is, however, not understood.

In this paper, we report the serendipitous discovery of
a long and narrow ($\sim$60 kpc$\times$2 kpc)
ionized region associated with a poststarburst galaxy D100
(\citealp{Dressler1980}, or GMP 2910, \citealp*{Godwin1983})
in the Coma cluster.
We assume that the distance modulus 
of the Coma cluster as $(m-M)_0=35.05$ and
($h_0$,$\Omega_M$,$\Omega_\lambda$)=(0.73,0.24,0.72) \citep{Spergel2006}.
Under these assumptions,
1 arcsec corresponds to 0.474 kpc at the distance.

\section{Imaging Observation and Result}

We observed 34$\times$27 arcmin region near the Coma cluster center
($\alpha,\delta$)(J2000.)=(12:59:26, +27:44:16)
with Suprime-Cam \citep{Miyazaki2002}
at the Subaru Telescope on 28 Apr 2006 and 3 May 2006 UT (Table 1).
We used three broad band filters,
$B$, $R$, $i$, and a narrow
band filter (N-A-L671, hereafter \NB).
The \NB\ filter is designed for observing \Ha\ emitting
objects in the Coma cluster at $z=0.0225$,
and has bell-shaped transmission with central wavelength of
6712 \AA\ and FWHM of 120 \AA.
The imaging data were reduced in standard manner.
The limiting surface brightness and the PSF sizes
of the final image are summarized in Table 1.
We adopted SA113 \citep{Landolt1992}
for $R$ and $B$ band photometric standard. 
The photometric zero points are converted to AB magnitude,
assuming AB-Landolt is 0.169 and -0.140 for $R$ and $B$,
respectively \citep{Fukugita1995}.
The flux of \NB\ and $i$ band are calibrated
with spectrophotometric standard;
GD153 \citep{Borlin1995} for \NB\ and HZ44 \citep{Oke1990} for $i$.
The magnitude of the stars are
m(GD153)=13.77(\NB) and m(HZ44)=12.35($i$) in AB system.

In order to create the \Ha\ image, we subtract scaled
$R$-band image from \NB\ image,
so that the residual of typical stars and galaxies should be minimal.
The resulting \Ha\ image is shown in Figure 1, together with
the broad band images.
A striking feature can be seen in the \Ha\ image:
a narrow and straight ionized region reminiscent of a jet
extends from the galaxy D100 to the north-east.
The contour of \NB\ surface brightness is shown in Figure 2.
D100 is an Irr or Sab galaxy of $M_r\sim-20$ magnitude.
It is noted that the direction of the major axis of D100
is slightly different between the \Ha\ image and three broad band
images.
Its neighbor D99 has different velocity,
and thought to be a chance overlap (\citealp*{Caldwell1999}; hereafter C99).
The third galaxy GMP 2913 
has no spectroscopic information in the literature.

We carefully surveyed archived data of D100,
and found that the feature is vaguely recognized
in the shallow $R$-band image
taken at the William Herschel Telescope (WHT)
in 1996 \citep{Komiyama2002}, and in the $R$-band Subaru
image\footnote{taken from Subaru data archive system SMOKA http://smoka.nao.ac.jp/}
taken in 2001. Bright parts are also visible in the shallow
\Ha\ image of Universidad Complutense de
Madrid survey \citep{PerezGonzalez2003}
\footnote{http://t-rex.fis.ucm.es/ucm\_survey/}.
The feature is therefore not transient
over a time scale of $\sim$10 years.

\section{Spectroscopic Observation and Result}

The spectra of some part of the region were
taken with Faint Object Camera and Spectrograph
\citep{Kashikawa2002} in multi object spectroscopy mode
on 23 Jun 2006 UT.
We used short slits whose lengths were 
6 -- 16 arcsec to obtain the spectra of bright 
filaments in the ionized region.
The width of the slits was $0\farcs8$, giving the spectral
resolution $R \sim$700 with 300B grism.
We obtained eight sets of 30 minutes exposures.
The sky condition was not photometric.
We observed HZ44 \citep{Oke1990} for relative flux calibration.

The data were reduced in standard manner;
bias subtraction, flat-fielding and distortion correction.
For wavelength calibration, we used
night-sky emissions. 
Background sky spectrum is constructed for each exposure
from spectra where no feature is detected
either in our \NB\ image or \Ha\ image.
The sky subtracted spectra of the eight exposures are
then scaled and coadded with 3-sigma clipping.
Examples of the resulting spectra are shown in Figure 3.

The redshift of D100 center (ID=01) is estimated to be $z=0.01784$
from \Ha, [NII] and [SII] emissions.
The value is consistent with $z=0.01776\pm0.00017$
of D100 itself reported by \citet{Smith2004}.
The vertical lines in Figure 3
indicate emission at the estimated redshift of D100.
We also obtained a spectrum of neighbor galaxy GMP 2913.
The galaxy shows strong \Ha\ absorption,
and the redshift estimated from the absorption is $z=0.0174$.
The proximity of the redshift and the projected distance
($\sim$9 kpc) suggests that GMP 2913 may have interacted, or
be interacting with D100.

Each spectrum of emission line region
is fitted with a linear continuum plus
three-gaussian (\Ha, and [NII]$\lambda\lambda$6548,6583) model
in 6600 \AA$<\lambda_{\rm obs}<$6800 \AA.
Since [SII]$\lambda\lambda$6717,6731 emission-lines 
at z=0.018 overlap
both atmospheric emission and absorption,
we did not include [SII] in our analysis.
We also fit a single gaussian+continuum model to
\Hb\ and [OIII]$\lambda$5007, separately.
The flux ratios of best-fit values are summarized in Table 2.
The relative radial velocity data (column 4 in Table 2)
show that the redshifts of ionized regions are
close to the redshift of D100, indicating that
the region is physically associated with D100.
Since the spectra were obtained under non-photometric condition,
we calculate \Ha\ flux (Table 2 column 8) from
\NB\ surface brightness of imaging data (Table 2 column 5),
\NB\ filter response and observed spectrum.

C99 described spatially resolved spectra of D100 in detail.
They reported that the galaxy has both strong emission and
strong underlying Balmer absorption in the central 2 arcsec
while only Balmer absorption is seen at 3 arcsec radius.
Such configuration is also seen in our Figure 1.
This suggests that central starburst was triggered shortly
after star formation in the disk stopped by some reason.
In our spectra, \Ha\ absorption is seen at 
the regions over the core.
In ID=02 spectra, which is in the emission line region,
\Hb\ emission is buried in strong Balmer absorption, 
and only [OIII] is detected.
We therefore regard that among the values in Table 2
[NII]/\Ha\ and [OIII]/\Hb\ is upper limit,
and \Ha\ brightness is lower limit
in the regions where continuum is detected (ID=01-04).

There is no signs of AGN activities because
the ratio [OIII]$\lambda 5007$/\Hb\ is smaller than 0.49 at the center.
The lack of AGN sign is consistent with previous studies of the galaxy
\citep[C99]{Quillen1999}.

\section{Characteristics of the extended region}

A distinct characteristic of the ionized region
is its morphology; it is narrow ($\sim$2 kpc),
long ($\sim$60 kpc), and straight.
The region ends at the central region of D100,
and no emission is found at the opposite side (Figure 1).
The velocity field of the region is distorted
and does not show any smooth global gradient,
but it suggests that the region is kinematically connected to
the center of D100 (Table 2).
The nucleus of D100 shows starburst characteristics and
its disk shows a poststarburst feature whose age is
$\sim$0.25 Gyr (C99).

\citet{BravoAlfaro2000} observed HI around the center of 
Coma with typical threshold of 2-4 $\times 10^{19}$ cm$^{-2}$.
Though their observed region include D100, 
they found no HI feature around D100.
Regarding X-ray, \citet{Finoguenov2004} presented resolved 
X-ray map of Coma cluster center. In the map, no feature
is recognized around D100.

The total mass of the ionized gas 
is estimated as follows.
Since we have no information about inclination, 
we assume that the extention of the region 
is perpendicular to the line of sight, and it is approximated 
as 60kpc$\times$1kpc radius cylinder, hereafter.
The typical \Ha\ surface brightness of the region is
$\sim 2\times10^{-17}$ erg cm$^{-2}$ s$^{-1}$arcsec$^{-2}$
(Table 2). 
It corresponds to \Ha\ photon flux of $3\times 10^{5}$
cm$^{-2}$ s$^{-1}$ str$^{-1}$.
The case B recombination coefficient at 
$T_e \sim 10^4$ K is 
$\alpha_B \sim 2.6 \times 10^{-13}$ cm$^3$ s$^{-1}$
\citep{BlandHawthorn1999}.
Assuming optically thin gas and isotropic radiation field,
the emission measure of the region is 
estimated as $n_e^2 L \sim 5 $ cm$^{-6}$ pc.
We assume that the thickness of the region $L \sim 2$ kpc, 
and then the rms of the electron density is
$\langle n_e \rangle \sim 0.05$ cm$^{-3}$.
If the gas is totally ionized and have uniform distribution,
the mass of the ionized gas is about
$\pi$ 1kpc$^2$ $\times$ 60kpc $\times$ 
$\langle n_e \rangle$ $\times$ $m_H$ $\sim$ 
2 $\times 10^8$ M$_{\odot}$.
The order of the estimated mass 
is comparable to the order of total gas mass in a dwarf galaxy.
If ionization fraction is lower, the value is underestimated.
However, if ionization fraction is lower than 0.9, for example,
the HI column density is $>$ 3 $\times 10^{19}$ cm$^{-2}$
and must have been detected in HI observation by \citet{BravoAlfaro2000}
whose threshold is 2-4 $\times 10^{19}$ cm$^{-2}$.
We can conclude that the ionization fraction should be almost unity.
If filling factor is low, on the other hand, the 
total mass should be smaller. The estimated value, 
$\sim$ 2 $\times 10^8$ M$_{\odot}$, is therefore the upper limit.

\section{Possible explanation}

At first glance, the straight morphology of the ionized region
gives us the impression that it is a jet ejected from the
galaxy nucleus.
The lack of current AGN activity in D100 and no detection of
radio jet around D100, however, make it
difficult to interpret the ionized region as an AGN jet.
In the following, we discuss on possible scenarios for
the origin of the extended ionized region and answer 
the questions,
1) the origin of the gas,
2) the mechanism to form the long and narrow morphology,
and 3) the ionization source.

\subsection{Infalling dwarf scenario}
One possibility is that the gas came from 
other object than D100; galaxy or gas cloud.
The object would possibly be dissipated or
absorbed by D100, because
we can not find any giant galaxies near the region.
The neighbor dwarf, GMP 2913, is apparently 
not connected to the region, and therefore 
GMP 2913 might not be the origin of the gas.
A possible candidate is 
a small clump seen  4 arcsec away from the center of D100
(corresponds to ID=03 spectra in Table 2 and Figure 2),
which could be a remnant core of infalling dwarf.

In this infalling dwarf hypothesis, 
either ram-pressure to the dwarf or 
tidal force by D100 can be a mechanism to form the region.
A famous example of an elongated gas cloud without stars
is MS. 
The width of MS is estimated to be $<$ 3.5 kpc (WW96). 
The length of MS can be calculated as $\sim$50 kpc
from its extension in the sky 
\citep[$\sim$50 degree; e.g.,][]{Bruns2005},
and the distance from us ($\sim$50 kpc).
The size is comparable to the extended region discussed here. 
The total HI mass of MS, 2-5$\times$10$^8$ M$_\odot$\citep{Moore1994,
Putman2003a,Bruns2005}, is also comparable to 
the estimated gas mass of the region.
The mechanism which formed MS is still under argument
\citep*[e.g.,][and references therein]{Cornors2006}, 
but the same mechanism might have worked here.
WW96 reported that the brightest parts of MS emit
$\sim$2$\times$$10^{-17}$erg cm$^{-2}$ s$^{-1}$
arcsec$^{-2}$ of \Ha\, which is comparable to the \Ha\ brightness
observed for the region in this study(see Table 2).
The difference is that 
the \Ha\ emitting region in MS is patchy \citep{Putman2003b},
while the region in this study has smooth and wide-spread 
\Ha\ emission (Figure 1,2).
The difference may be explained by difference of ionization source.

The ionization source of bright spots of MS 
is first thought to be a friction with 
hot plasma of Galactic halo (WW96).
Though the mechanism is argued to be insufficient 
for the luminosity of the bright spots in MS
\citep{BlandHawthorn1999,BlandHawthorn2001,BlandHawthorn2002},
it could work effectively in hot intracluster plasma 
in Coma cluster.
\citet{Putman2003b} suggested that 
some bright spots in MS, which are an order of magnitude stronger
in \Ha, may be produced through the interaction with halo debris.
Similarly, the interaction with surrounding plasma in the Coma cluster 
could ionize the region in this study.
We can see enhanced [NII]/\Ha\ and [OIII]/\Hb\ ratios
at some part of the extended emission region,
which implies a moderately low energy shocks of
moderately metal rich gas.

To explain the \Ha\ emission of MS,
\citet{BlandHawthorn1999} showed that
escaping photons from Galactic disk can ionize the spots.
We checked whether escaping photons 
from D100 can ionize the region using simple model.
The apparent size of \Ha\ emitting region of D100 is 
$\sim 500$ arcsec$^2$=1.2$\times 10^{-8}$ str.
Since typical \Ha\ photon flux density of the extended region is 
$3\times 10^{5}$ cm$^{-2}$ s$^{-1}$ str$^{-1}$,
the total \Ha\ photon flux is $3.5\times 10^{-3}$cm$^{-2}$ s$^{-1}$. 
Assuming the distance to the region from us 
to be 100Mpc,
total \Ha\ photon flux from the region is 
4$\times$10$^{51}$ s$^{-1}$  if the region is cylindrical.
Since most of the disk is now in poststarburst phase,
we simplify the model that ionizing photons are created 
only at the core, and calculate 
\Ha\ photon from a cone of 60kpc depth and 1kpc radius circle,
whose opening angle is 8.7$\times$10$^{-4}$ str.
Since the volume of cone is 1/3 of cylinder,
the flux from the cone is 1.3 $\times$10$^{51}$ s$^{-1}$.
Assuming spherical symmetry,
the whole ionizing photon flux from the core is 
estimated as 1.9$\times 10^{55}$ s$^{-1}$.
Following \citet{Kennicutt1998},
the starforming rate required to ionize the region 
is $\sim$ 200 M$_\odot$ yr$^{-1}$ in D100.
If there has been a strong starburst at the D100 core
for $\sim$ 2$\times 10^5$ years (60kpc / light speed),
and the core region is free from dusts, 
the photon from the core can ionize the whole extended region.

Yet another ionizing source is 
EUV photons from hot plasma \citep{Maloney2001}.
\citet{Maloney2001}
discussed that EUV from hot gas in cluster of galaxies
can be ionizing source of gas clouds.
Such ionizing EUV is found to be strong 
in Coma cluster \citep{Bowyer2004},
and it is also possible that
the extended region is ionized by the EUV.
Any of the three possible sources and their combination
can ionize the extended region.

A possible infalling dwarf scenario is as follows,
(1) A dwarf galaxy was trapped by the gravity of D100
and started to interact. 
(2) The gas in D100 disk lost angular momentum
and fell into core. This stopped star formation 
in the disk and central starburst is triggered.
(3) The dwarf expelled gas as a stream
by the same mechanism as MS was formed.
(4) The gas is orbiting around D100, or infalling into D100
and fully ionized by some mechanism discussed above. 

The problem of this scenario is the configuration of the region.
The stream seems to be smoothly connected to 
the core and there is no feature in the
other side of the core.
To reproduce such an appearance, the stream should end or 
be truncated just at our line of sight to D100.
Moreover, as we see a straight morphology,
the stream is observed as edge-on.
The probability of such a configuration is very low.

The absense of smooth velocity gradient 
is another difficulty, since such stream is thought to have 
smooth velocity gradient.

\subsection{Ram pressure stripping scenario}

We should consider the other origin of the gas;
the gas came from the disk of D100. 
Under this assumption, the mechanism to form the region would 
be the ram pressure stripping by surrounding gas.
C99 noted that D100 shows little rotation at
121$^\circ$ position angle, which
might imply that D100 have shallow gravitational potential.
If this is the case, it is possible to strip the central gas of
D100 very far away from the galaxy by ram pressure.
Two examples of ionized gas stripped by ram pressure
are reported by \citet{Gavazzi2001} in Abell 1367,
whose sizes are 75$\times$8kpc and 50$\times$8kpc.
Another example is discovered by \citet{Sun2006} in A3627
as 71$\times$8 kpc X-ray tail.
The ionizing source of the region can be the same as that of the 
infalling dwarf scenario; escaping photon from D100,
moderate internal shock, and/or EUV photon from hot plasma.

In this scenario, the formation of the region is as follows.
(1) A dwarf galaxy was trapped by the gravity of D100
and started to interact. The 
interacting galaxy could be neighboring GMP 2913
or a small clump near the core (ID=03).
(2) The gas in D100 disk lost angular momentum
and fell into core. This stopped star formation in the disk and
central starburst was triggered.
(3) The condensed central gas received ram-pressure of
intracluster medium (ICM) and is stripped
with disturbed velocity.
(4) The gas was blown far out of the disk.
This scenario explains the distorted velocity field in the region
and the fact that the end of the stream 
is connected to the core.

A difficulty of the scenario is that 
the ionized region is too narrow and straight 
(2$\times$60kpc) compared with 
the examples found in previous studies 
(e.g., 75$\times$8kpc, 50$\times$8kpc \citealp{Gavazzi2001})
and simulations \citep[e.g.][]{Roediger2006}.
\citet{BlandHawthorn1995} discussed 
that intracluster plasma can confine ionized gas
in Fornax cluster.
Similar confinement process 
may be able to confine such a narrow ($<2$kpc) and 
long extended region in Coma cluster, 
since surrounding hot plasma is much denser than Fornax.
This scenario also requires some reason why ram pressure began 
to act just after D100 experienced the merger.
If the infall of disk gas was induced not by the minor merger,
but by the ram pressure, the coincidence is explained in natural.
Though it is known that most of field poststarburst galaxies are
created by merger/interaction \citep[e.g.][and references therein]
{Zabludoff1996, Blake2004, Goto2005},
such ram-pressure induced poststarburst in rich clusters
has also been suggested by some previous studies 
\citep{Poggianti2004,Pracy2005}.
Currently we do not have simulations to reproduce
such a morphology of ionized gas.

We will need some additional data,
such as resolved and much deep X-ray and radio data,
metallicity of the gas, to investigate 
the nature of the region in detail.
As we set slit along the region,
spectroscopy of higher spatial resolution 
across the region would give us another hints
about velocity structure and gas excitation,
which may set 
constraints on gas properties and 
confinement mechanism.
We also require some model simulations to reproduce 
the morphology.

\acknowledgments

We are sincerely grateful to the referee Prof. Bland-Hawthorn 
for his thorough reading and many suggestive comments
which improved this paper very much. 
This work is based on data collected at Subaru Telescope,
which is operated by NAOJ. We appreciate Subaru staffs for their help.
This work has made use of NED database\footnote{http://nedwww.ipac.caltech.edu/}, SMOKA archive and sb-system at Astronomical Data Center of NAOJ.

\clearpage

\begin{table}
\begin{tabular}{|c|c|c|c|c|}
\hline
filter &
PSF size & SB$_{\rm lim}$(ABmag arcsec$^{-2}$) &
Date(UT) & exposure \\
\hline
$B$ & 1''.06 & 28.8 & 2006-04-28 &  2$\times$450sec  \\
    &        &      & 2006-05-03 &  3$\times$450sec  \\
\hline
$R$ & 0''.75 & 27.7 & 2006-04-28 & 11$\times$300 sec  \\
    &        &      & 2006-05-03 &  1$\times$60 sec  \\
\hline
$i$ & 1''.10 & 27.2 & 2006-04-28 & 5$\times$240 sec  \\
    &        &      & 2006-05-03 & 2$\times$240 sec + 1$\times$120 sec 
\\
\hline
\NB & 0''.76 & 27.7 & 2006-04-28 &  8$\times$1800 sec \\
    &        &      & 2006-05-03 &  3$\times$120 sec   \\
\hline
\end{tabular}
\caption{Summary of mosaicked images and observation log.
SB$_{\rm lim}$ represent 1-$\sigma$ fluctuation of
1 arcsec $\times$ 1 arcsec regions}
\end{table}

\begin{table}
\begin{tabular}{|c|c|c|c|c|c|c|c|c|}
\hline
ID &
Length&
Distance&
$V_{H\alpha}$&
SB$_{\rm\NB}$&
[NII]/\Ha &
[OIII]/\Hb &
F$_{H\alpha}$ \\
(1) & (2) & (3) & (4) & (5) & (6) & (7) & (8) \\
\hline
01  & 2.7 &   0.0 &  0   & 18.5 & $<$0.47 & $<$0.45 & $>$233.\\
02  & 1.2 &   2.2 & +24  & 20.4 & $<$0.86 & [OIII]\tablenotemark{b}  & 
$>$10.2\\
03  & 1.2 &   3.8 & +59  & 21.3 & $<$0.53 & $<$0.45 & $>$7.0 \\
04  & 1.2 &   8.2 & +122 & 23.7 & $<$0.85 & $<$0.59 & $>$1.8 \\
05  & 1.5 &  11.7 & +55  & 24.4 & 0.86   & --      & 1.2   \\
06  & 1.2 &  16.3 & +71  & 24.8 & 0.57   & --      & 1.5   \\
07  & 1.2 &  18.1 & +105 & 24.6 & 0.80   & --      & 1.3   \\
08  & 1.5 &  20.6 & +63  & 24.2 & 0.52   & --      & 2.3   \\
09  & 2.4 &  23.9 & +116 & 24.2 & 0.73   & 0.67    & 2.1   \\
10  & 0.9 &  31.1 & +314 & 24.2 & 0.64   & \Hb\tablenotemark{c} & 2.6   \\
11  & 1.5 &  32.3 & +78  & 24.5 & 0.66   & \Hb\tablenotemark{c}      & 2.1 
\\
12  & 2.1 &  40.9 & +74  & 24.6 & 0.87   & 0.58    & 1.9   \\
13  & 1.8 &  50.3 & +150 & 24.9 & 0.63   & \Hb\tablenotemark{c}      & 1.8 
\\
14  & 2.1 &  53.1 & +134 & 25.1 & 0.61   & --      & 1.3   \\
15  & 2.7 &  60.4 & +161 & 25.8 & 0.65   & --      & 0.64  \\
16  & 1.5 &  64.3 & +2   & 24.8 & 0.57   & \Hb\tablenotemark{c}      & 1.3 
\\
17  & 2.1 &  66.1 & +110 & 23.9 & 0.44   & \Hb\tablenotemark{c}      & 3.7 
\\
18  & 1.5 &  89.3 & +182 & 25.1 & 0.59   & --      & 1.3   \\
19  & 2.4 &  91.2 & +130 & 24.3 & 0.54   & 0.29    & 3.1   \\
20  & 2.1 &  94.0 & +138 & 24.9 & 0.81   & 0.42    & 1.5   \\
21  & 1.2 & 118.0 & +303 & 26.0 & 0.88   & --      & 0.64  \\
22  & 3.0 & 121.3 & +149 & 26.2 & 0.69   & --      & 0.59  \\
\hline
\end{tabular}
\caption{Summary of spectral features\tablenotemark{a}}
\tablenotetext{a}{
(1) aperture ID,
(2) integration length along slit [arcsec],
(3) projected distance from the D100 center [arcsec],
(4) radial velocity [km s$^{-1}$] relative to D100 center,
(5) \NB\ surface brightness [AB mag arcsec$^{-2}$],
(6) flux ratio of [NII]$\lambda\lambda 6584$ to \Ha,
(7) flux ratio of [OIII]$\lambda\lambda 5007$ to \Hb,
[OIII] or \Hb\ means only the line can be fitted,
and
(8) estimated H$\alpha$ flux [10$^{-17}$ erg s$^{-1}$ cm$^{-2}$
arcsec$^{-2}$].}
\tablenotetext{b}{Only [OIII] is detected}
\tablenotetext{c}{Only \Hb\ is detected}
\end{table}

\begin{figure}
\includegraphics[scale=0.55]{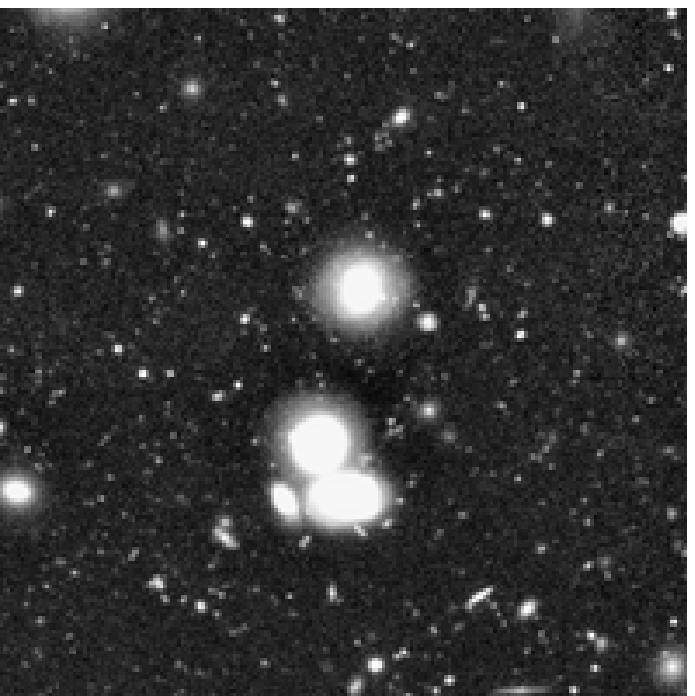}
\includegraphics[scale=0.55]{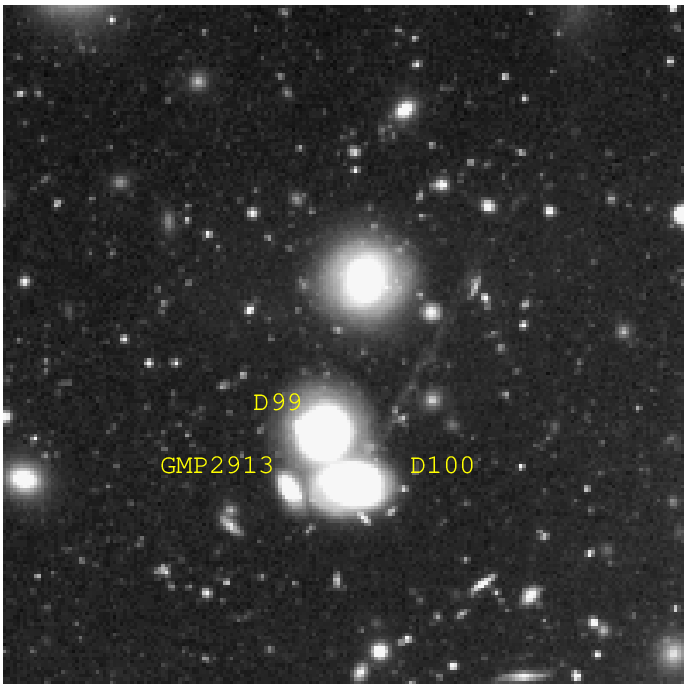}
\includegraphics[scale=0.55]{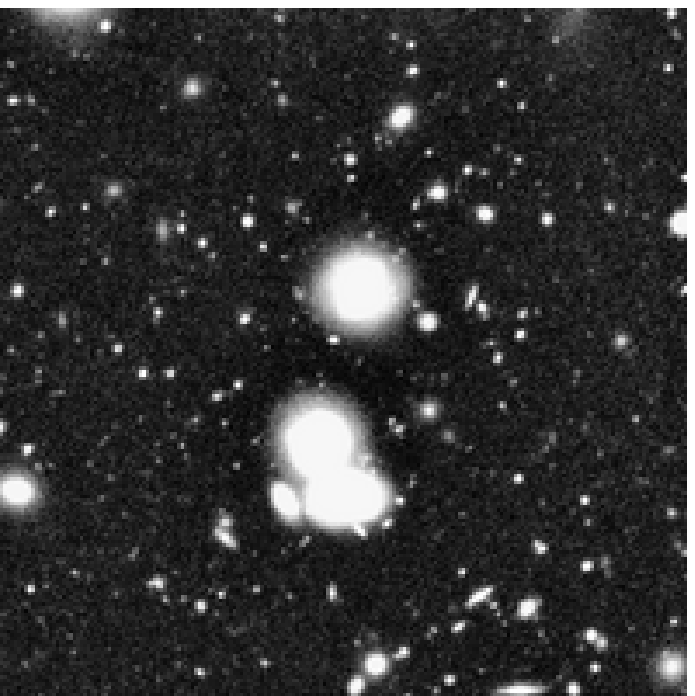}
\includegraphics[scale=0.55]{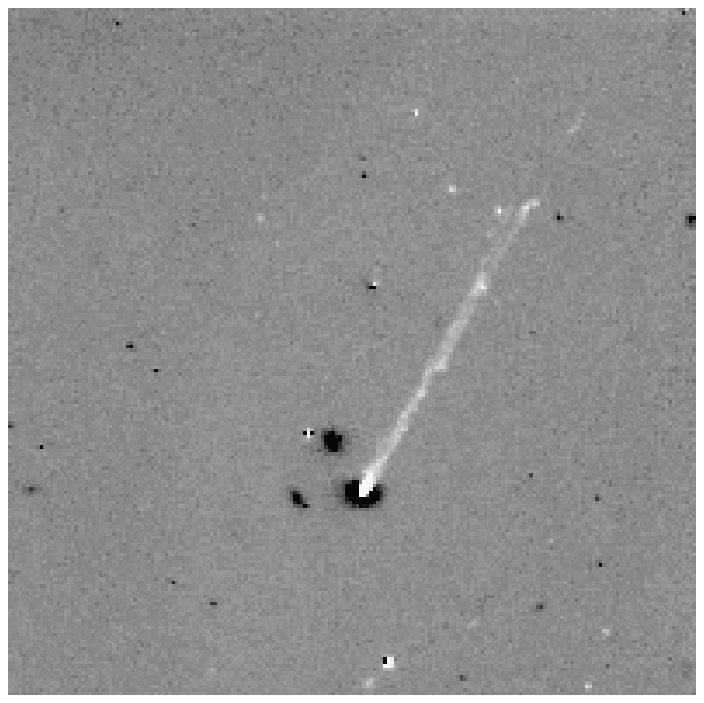}
\caption{
Images around Mrk 60 (D99+D100 system) in
$B$, $R$, $i$ and \Ha\ (=$NB-R$) from the left to the right.
The size is 200$\times$200 arcsec and east is up.
In \Ha\ image, white represents \Ha\ emission and black
represents \Ha\ absorption.
}
\end{figure}

\begin{figure}
\includegraphics[scale=0.5]{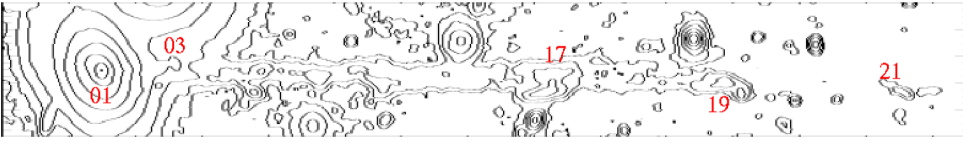}
\caption{\NB\ surface brightness contour.
The faintest isophote is 27 mag arcsec$^{-2}$,
and the interval is 1 mag arcsec$^{-2}$.
Numbers indicate aperture ID in Table 2,
for apertures shown in Figure 3, and possible merger remnant (ID=03).
}
\end{figure}

\begin{figure}
\includegraphics[scale=0.3,angle=-90]{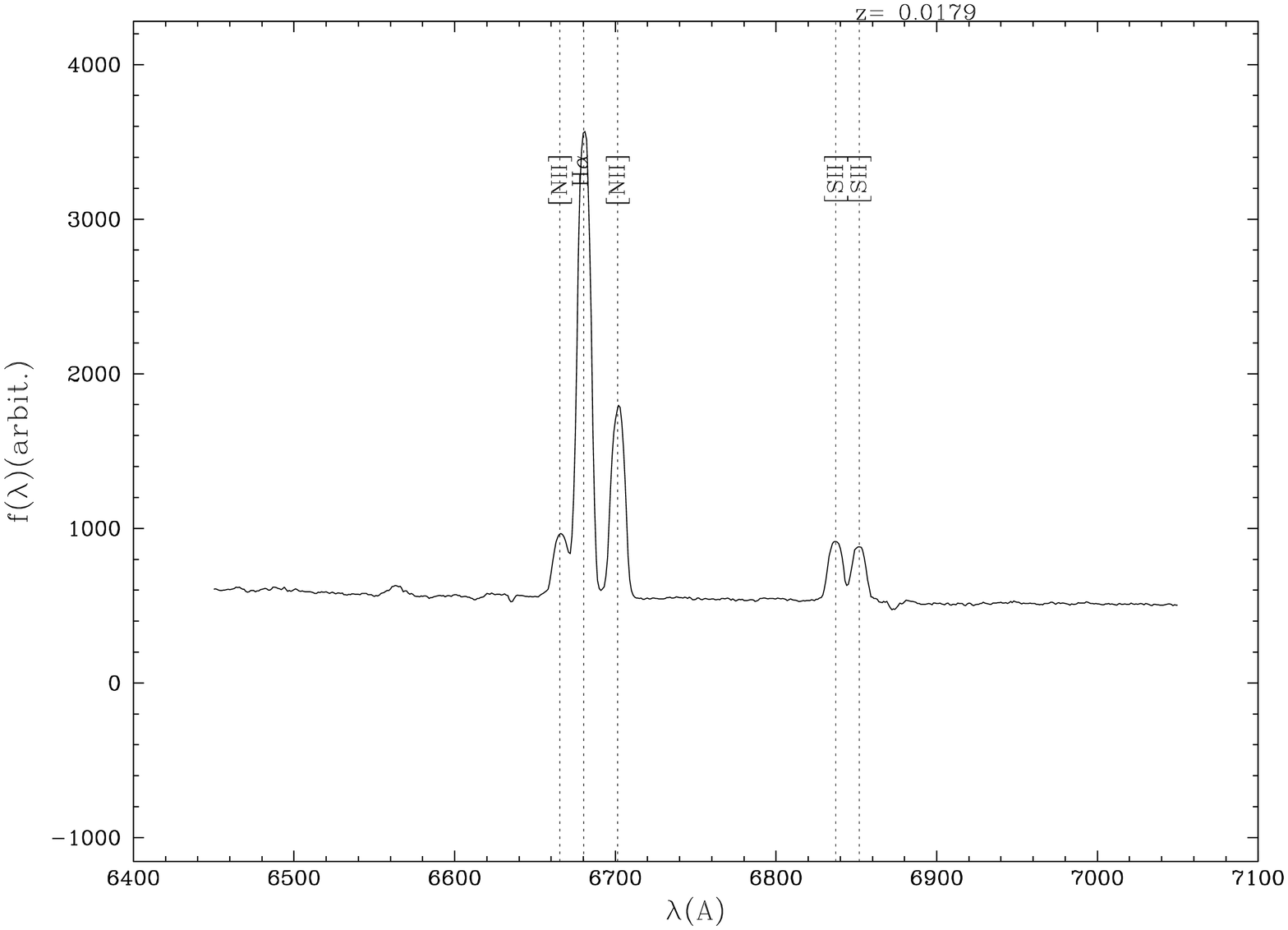}
\includegraphics[scale=0.3,angle=-90]{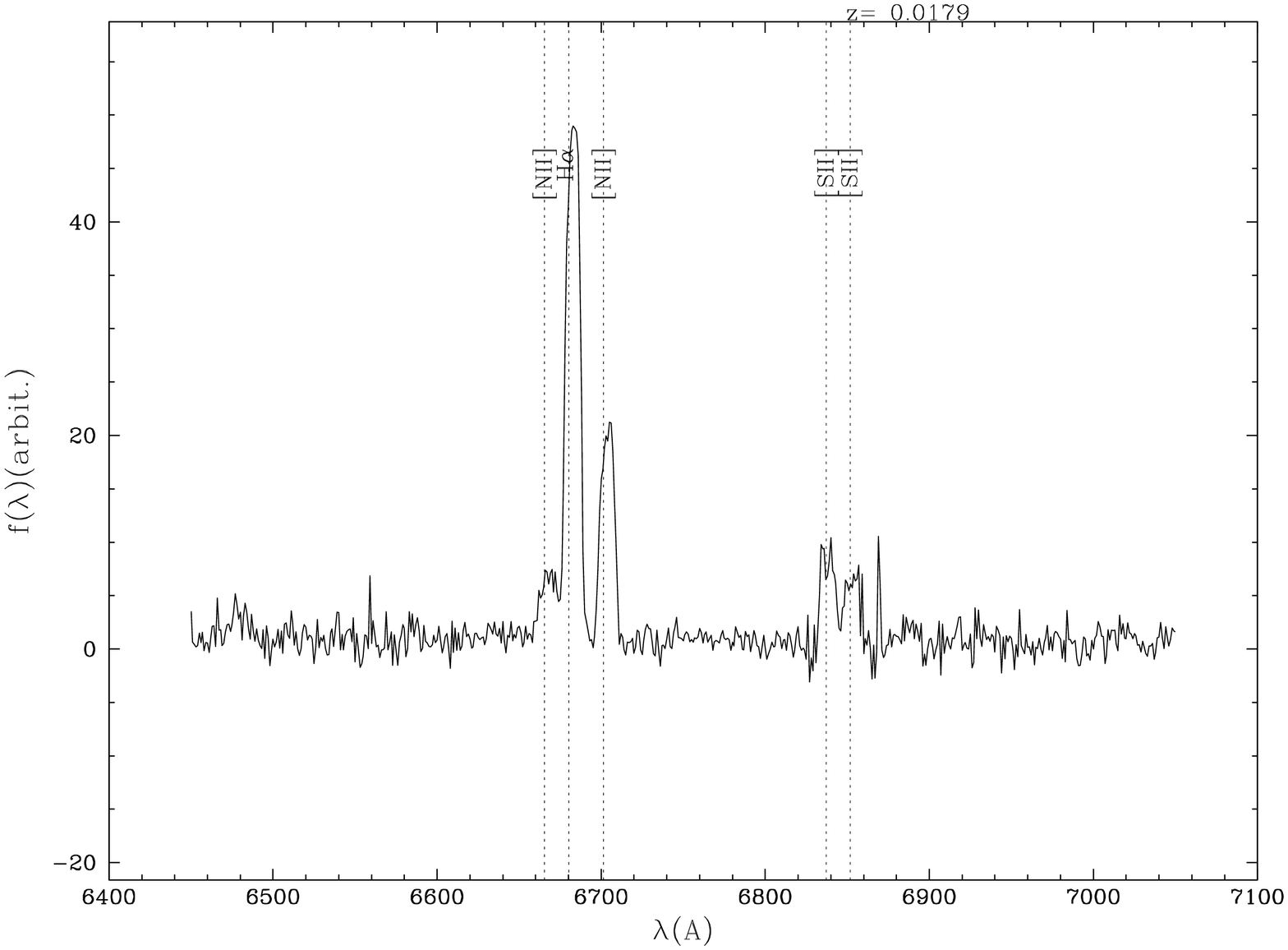}\\
\includegraphics[scale=0.3,angle=-90]{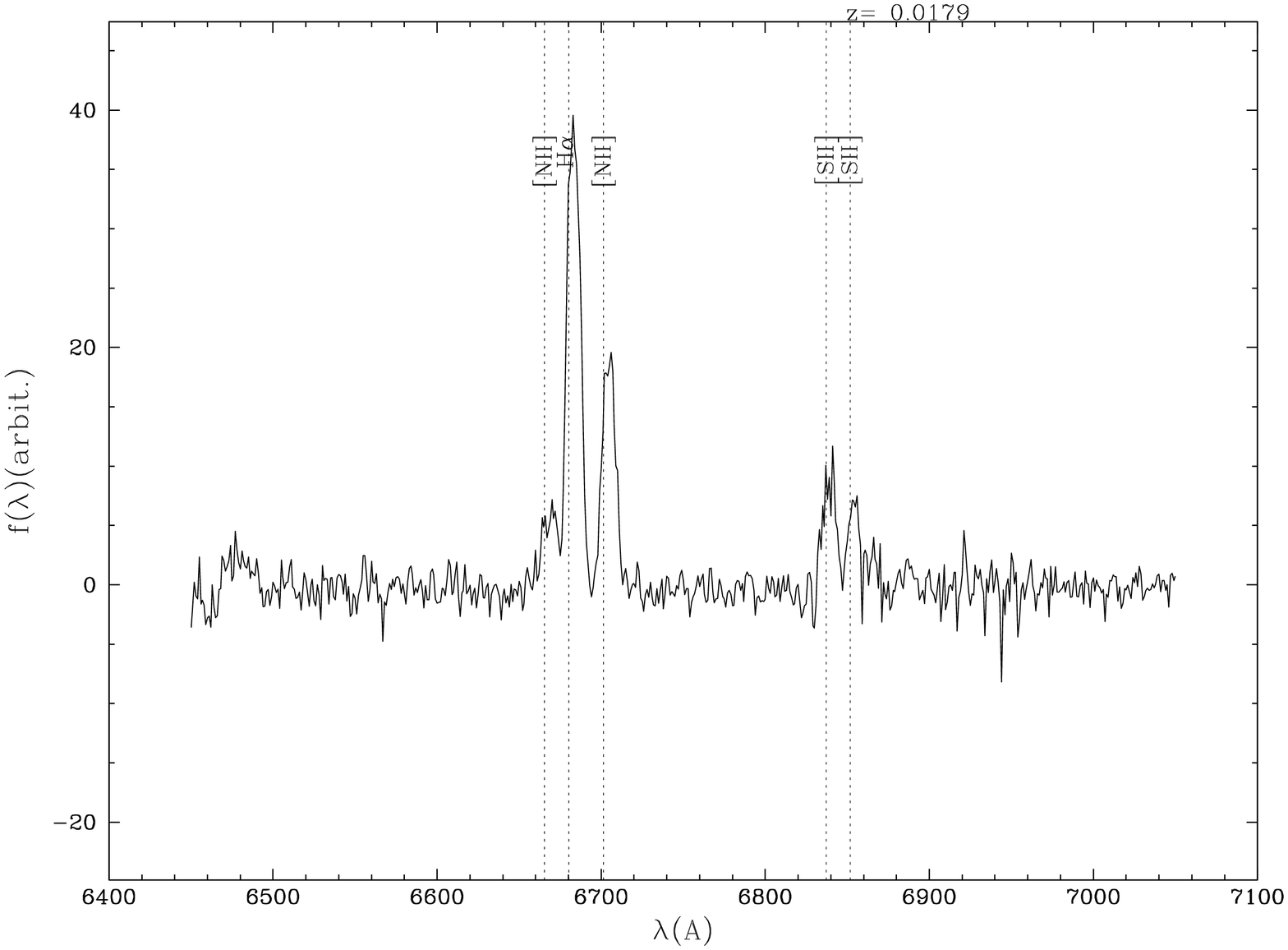}
\includegraphics[scale=0.3,angle=-90]{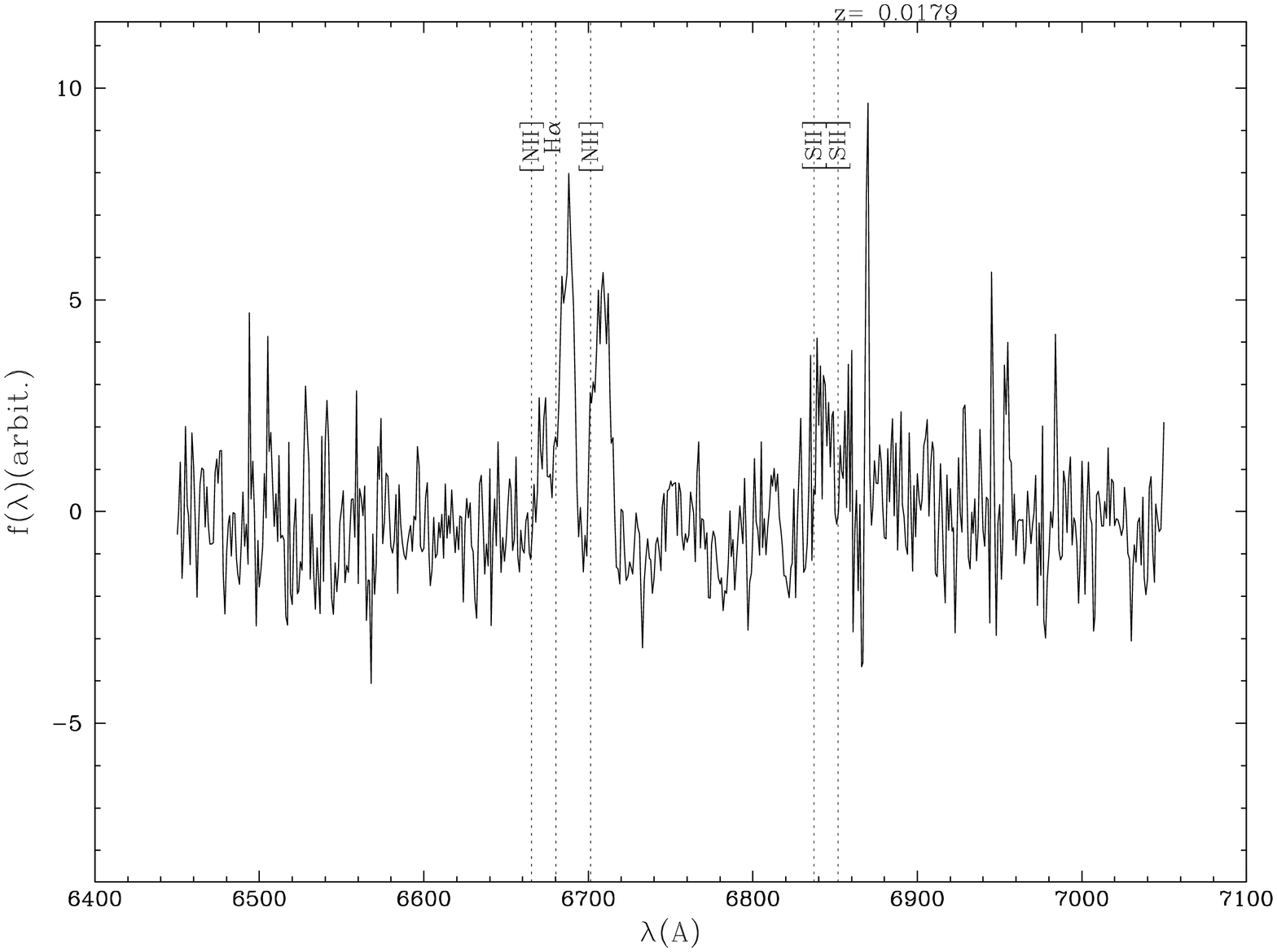}\\
\caption{The spectra of D100 center (top-left, ID=01 in Table 2),
and bright spot in the feature (ID=17 top-right, ID=19 bottom-left,
and ID=21 bottom-right).
The vertical lines indicate the best-fit 
wavelength of  emission lines determined at the ceter (ID=01).
}
\end{figure}

\end{document}